\documentclass{ws-mpla}

\begin{document}

\markboth{T. Rindler-Daller, P.R. Shapiro} {Complex scalar field dark
matter on galactic scales}

%%%%%%%%%%%%%%%%%%%%% Publisher's Area please ignore %%%%%%%%%%%%%%
\catchline{}{}{}{}{}
%%%%%%%%%%%%%%%%%%%%%%%%%%%%%%%%%%%%%%%%%%%%%%%%%%%%%%%%%%%%%%%%%%%

\title{COMPLEX SCALAR FIELD DARK MATTER \\ON GALACTIC SCALES}

\author{\footnotesize TANJA RINDLER-DALLER$^{1,2}$ and PAUL R. SHAPIRO$^{2}$}

\address{$^{1}$Department of Physics and Michigan Center for Theoretical Physics, University of Michigan,\\
450 Church Street, Ann Arbor, MI 48109, USA}

\address{$^{2}$Department of Astronomy and Texas Cosmology Center, The University of Texas at Austin,\\
2515 Speedway C1400, Austin, TX 78712, USA}

\maketitle

\pub{Received (Day Month Year)}{Revised (Day Month Year)}

\begin{abstract}

The nature of the cosmological dark matter remains elusive. Recent
studies have advocated the possibility that dark matter could be
composed of ultra-light, self-interacting bosons, forming a Bose-Einstein condensate
in the very early Universe. We consider models which are charged
under a global $U(1)$-symmetry such that the dark matter number is conserved. 
It can then be described as a classical complex scalar
field which evolves in an expanding Universe. We present a brief
review on the bounds on the model parameters from cosmological and
galactic observations, along with the properties of galactic halos
which result from such a dark matter candidate.

\end{abstract}

\ccode{PACS Nos.: 98.80.-k; 95.35.+d; 98.62.Hr; 14.80.-j}

\section{Introduction and Motivation}

The nature of the cosmological dark matter is one of the most
profound open questions in modern physics and cosmology. After many
decades of research, it has become clear that dark matter underlies
the formation of structure as we see it in the Universe today:
galaxies and galaxy clusters reside in high-density filaments which
surround voids where the density is comparatively low, giving rise
to the large-scale cosmic web of structure. Numerical cosmological
simulations have shown that large-scale structure is best
represented if the dark matter is assumed to be a collisionless and
cold, i.e. non-relativistic, entity. On cosmological scales,
therefore, it behaves like a 'dust-like' fluid.

Meanwhile, theories beyond the standard model (SM) of particle
physics have been devised which are able to provide candidate
particles for the dark matter (DM). The most popular and possibly
best motivated candidates are the lightest supersymmetric particles
in supersymmetric extensions of the SM. These are
weakly-interacting (i.e. subject to the weak force), massive
particles (WIMPs). While models allow generic values for their mass between around
1 GeV to 10 TeV, direct and indirect detection
experiments, including accelerator searches, are inconclusive and partly contradictory on their preferred
exclusion limits. However, steady progress has been made in attempts to detect WIMP dark matter
and it seems conceivable that it will be detected within the next
decade, if it exists. However, in case detection experiments
continue to deliver null-results, this may hint to an entirely
different type of DM, e.g. the possibility that DM is a very low-energy 
phenomenon, behaving wave-like on macroscopic scales, instead
of particle-like. Such a form of DM will be the topic of this review article.

In fact, one such DM candidate, equally prominent to WIMPs, has
been the QCD axion, the pseudo Nambu-Goldstone boson (PNGB) that
arises in the dynamical solution of the CP problem of the strong
force (see Ref.\cite{Peccei} for a review). Indeed, with a mass of
around $10^{-5}$ eV and the fact that it is born non-thermally, the
QCD axion can Bose-Einstein-condense, exhibiting coherence on the
order of a de-Broglie wavelength of about $\lambda_{deB} \sim 186$ m
in a galactic halo with a virial velocity of $200$ km/s. However,
this $\lambda_{deB}$ is still a tiny number compared to galactic
scales, and the QCD axion behaves like collisionless, cold dark
matter (CDM) on all galactic and cosmological scales of interest in
the present Universe. However, it has been pointed out in
Ref.\cite{SY} that the subsequent thermalization of the QCD axion to
find a new ground state in the expanding background Universe may be
'memorized' such as to imprint different characteristics in the
galactic dynamics, as compared to standard cold DM (CDM). This would be an
interesting way of distinguishing WIMPs and axions by astronomical
means, after all.

However, we will not consider the QCD axion in this review, but
rather even much lighter particles, which are guaranteed to behave
quantum-mechanically and hence distinctively from CDM on galactic
scales. These DM candidates are motivated from a fundamental, as
well as from an astrophysical point of view, as follows. In the very
early Universe, PNGBs can arise generally when a global symmetry is
spontaneously broken, while non-perturbative effects on lower energy
scales break the symmetry explicitly, generating the (ultralow)
mass. Examples include the aforementioned QCD axion, as well as
familons, Majorons and related objects. Indeed, since theories
beyond the standard model involve new symmetries, many of them
global, which, upon breaking, will result in scalar and pseudoscalar
PNGBs, the interest in those as dark matter candidates has lately
seen a huge rise, see e.g. Ref.\cite{FHSW}, Ref.\cite{axiverse} and
references therein. Ultralight scalar fields can also result as
gravitational excitons in multidimensional cosmological models,
giving rise to the dark matter in our observable Universe, see e.g.
Ref.\cite{GZ}. The choice of a potential in the Lagrangian which
describes the scalar field will determine its cosmic evolution. We
will be focusing on a scenario in which an ultra-light boson of mass
around $10^{-21}$ eV/$c^2$ with a Higgs-like potential is
responsible for all of the dark matter in the Universe. The boson
shall be ``charged'' under a $U(1)$-symmetry, such that the dark
matter number is conserved over the entire cosmic evolution, once it
is in its condensed state, in which it will enter in the very early
Universe. Its effective Langrangian shall thus be given by
\begin{equation} \label{lag}
    \mathcal{L}=\frac{\hbar^2}{2m}g^{\mu\nu}\partial_\mu\psi^*\partial_\nu\psi - \frac{1}{2}mc^2|\psi|^2 - \frac{g}{2}|\psi|^4
\end{equation}
(with metric signature $(+,-,-,-)$). The quartic term describes an
effective 2-boson self-interaction, which we choose to be
non-negative\footnote{An attractive coupling leads to perpetual
instabilities and collapse, according to Ref.\cite{FMT} and
Ref.\cite{FM}.}, i.e. $g \geq 0$. As we will see, even tiny values
for the self-interaction coupling strength $g$ can render those
models surprisingly different from non-interacting ones.
Equ.(\ref{lag}) assumes no effective coupling of the scalar field $\psi$ to
SM particles, but only its minimal coupling to gravity\footnote{See
also Ref.\cite{Carroll}. On the other hand, Ref.\cite{FG} studies the
implications of a coupling to baryons for the cosmic microwave
background.}.

Scalar field dark matter (SFDM) provides a natural minimum scale for
gravitational equilibrium, once perturbations grow nonlinear in the
matter-dominated epoch. This makes it interesting for
astrophysicists, who have found that the predictions of structure
formation simulations using collisionless CDM are at odds with
observations on small scales, especially at the level of dwarf and
low surface-brightness galaxies. Simulations predict cuspy galactic
centers with DM densities going as $r^{-1}$ for $r \to 0$. In
addition, many hundreds of subhalos are expected to surround host
halos of Milky-Way size. On the other hand, observations tend to be
better fit by cored profiles, and the number of observed satellite
galaxies is smaller than predicted. These mismatches have been
around for two decades and are known as the cusp/core and 'missing
satellites' problems. In addition, it has been lately pointed out
that the known classical dwarf spheroidal satellite galaxies of the Milky Way are not
dense enough to populate the corresponding most massive subhalos
found in simulations, a problem called 'too-big-to-fail'. It appears
that all those problems could be cured if the DM densities in the
innermost parts of galaxies were lower. While CDM-only simulations
have their limitations and baryonic feedback processes have been
suggested to reconcile observations with simulations, it remains a
striking fact that it is just the DM-dominated galaxies which keep
challenging the CDM predictions most (see Ref.\cite{Weinberg} for a
recent review on these problems). The above-mentioned minimum scale,
therefore, is a welcoming feature of SFDM. That scale, however, is
determined by the boson parameters, mass $m$ and self-interaction coupling strength $g$. So, astronomical
observations can thereby help to establish or to rule out
high-energy extensions of the SM.

In this brief review, we will almost exclusively restrict to results
obtained for complex scalar fields with Lagrangian in
Equ.(\ref{lag}), but refer interested readers to Ref.\cite{MM} for a
review on real scalar fields and Ref.\cite{Guzman} for a wider
application of complex scalar fields. Also, we will only consider
works in which the scalar field itself constitutes the dark matter.

This article is organized as follows: in Section 2, we present basic
equations and results which we will need for our discussion of
galactic properties. Section 3 summarizes some recent results which
led to new bounds on the particle parameters $m$ and $g$. Finally,
Section 4 will address the implications of SFDM in general and of
these new bounds in particular on the structure of SFDM galactic
halos.

\section{Basic equations for complex SFDM in the late Universe}

\subsection{Equations of motion}

SFDM obeys the Klein-Gordon equation in an expanding Universe. As
soon as the rest-mass term dominates over other terms in the
Langrangian (e.g. the quartic term in Equ.(\ref{lag})), once the density
of SFDM drops significantly as a result of the expansion of the
Universe, the (background) equation of state of SFDM will be
dust-like. Since we assume that SFDM accounts for all of the dark
matter, SFDM gives then rise to the epoch of matter-domination. We
refer to this regime in the evolution of SFDM as Bose-Einstein condensed cold DM (BEC-CDM). In this
epoch, it can then be treated in a Newtonian way, with the
Klein-Gordon and Einstein equations of motion reducing to the
non-linear Schr\"odinger equation and the Poisson equation. This
holds true especially for the formation and evolution of galactic
halos, whose dynamics involves non-relativistic velocities and
gravitational fields\footnote{The only exception is the environment
close to a central galactic black hole, which we will not
consider.}. Thus, we will use the Newtonian framework of BEC-CDM in
our description of galactic halos. Nevertheless, the quartic term in
(\ref{lag}) is important in the early Universe and affects structure
on small scales in the late Universe, i.e. in the matter-dominated
epoch, as will be exemplified below.

The complex scalar field $\psi(\mathbf{r},t)$ describing the ground
state of SFDM in the matter-dominated era (BEC-CDM) satisfies the
Schr\"odinger-Poisson (also called Gross-Pitaevskii-Poisson (GPP)) system of
equations,
\begin{equation} \label{gp}
 i\hbar \frac{\partial \psi}{\partial t} = -\frac{\hbar^2}{2m}\Delta \psi + m\Phi \psi +
 g|\psi|^2\psi,
 \end{equation}
 \begin{equation} \label{poisson}
  \Delta \Phi = 4\pi G m |\psi|^2.
   \end{equation}
The number and mass density of DM in a given halo with gravitational
potential $\Phi(\mathbf{r},t)$ is then $n(\mathbf{r}) =
|\psi|^2(\mathbf{r})$ and $\rho(\mathbf{r}) = m n(\mathbf{r})$,
respectively.

The structure and evolution of BEC-CDM halos is governed by
quantum-kinetic energy, gravity, and the self-interaction of
identical dark matter bosons, as described by the quartic term in Equ.(\ref{lag}). The scattering cross section of
indistinguishable bosons becomes constant in the low-energy limit,
\begin{equation} \label{sigma}
  \sigma_s = 8\pi a_s^2
  \end{equation}
with the s-wave scattering length $a_s$. The coupling constant of
the effective interaction is then simply proportional to $a_s$,
\begin{equation} \label{coupling}
  g = 4\pi \hbar^2 a_s/m,
   \end{equation}
    which is the (first) Born approximation.

It is remarkable that Equ.(\ref{gp}) lends itself to a hydrodynamic
formulation, as originally observed by Ref.\cite{madelung}, an
attractive feature in the astrophysical context in which we use
(\ref{gp}). Inserting the decomposition of the complex field into modulus and phase,
  \begin{equation} \label{polar}
\psi(\mathbf{r},t) = |\psi|(\mathbf{r},t)e^{iS(\mathbf{r},t)} =
\sqrt{\rho(\mathbf{r},t)/m}~e^{iS(\mathbf{r},t)},
 \end{equation}
 into Equ.(\ref{gp}), leads to its splitting into a momentum and continuity equation,
 respectively,
  \begin{equation} \label{fluid}
     \rho \frac{\partial \mathbf{v}}{\partial t} + \rho (\mathbf{v} \cdot \nabla)\mathbf{v} = -\rho \nabla
     Q
    - \rho \nabla \Phi - \nabla P_{SI}
     \end{equation}
     and
 \begin{equation} \label{hd3}
\frac{\partial \rho}{\partial t} + \nabla \cdot (\rho \mathbf{v}) =
0,
 \end{equation}
   with the bulk velocity defined as $\mathbf{v} = \hbar \nabla S/m$.
The more intuitive representation of those equations as hydrodynamic
ones, however, comes at the expense of a more complicated,
higher-order derivative in the first term of the rhs of
(\ref{fluid}). In fact, the gradient of
  \begin{equation} \label{qpot}
   Q = -\hbar^2 \Delta \sqrt{\rho}/(2m^2 \sqrt{\rho})
    \end{equation}
  gives rise to what is often called 'quantum
 pressure', an additional force on the rhs of Equ.(\ref{fluid}), which basically stems from the quantum-mechanical
 uncertainty principle. It is responsible for the de Broglie length of the bosons as one characteristic length in the system:
 with the dimensions of $\Delta$ being L$^{-2}$ and changing to the momentum
 representation, we see that
  \begin{equation} \label{deB}
 L \sim \lambda_{deB} = h/p = h/(mv).
  \end{equation}
 On the other hand, the particle self-interaction results in a pressure
 of polytropic form in equ.(\ref{fluid}),
  \begin{equation} \label{selfpressure}
 P_{SI} = K\rho^2 \equiv  g/(2m^2) \rho^2.
  \end{equation}
We will see that the corresponding length scale is proportional to
$\sqrt{g/m^2}$ (see Equ.(\ref{onesphere})).

The system of equations (\ref{gp}) and (\ref{poisson}), or
(\ref{fluid})-(\ref{hd3}) with $\Delta \Phi = 4\pi G \rho$,
respectively, will determine the properties of BEC-CDM structures
which have decoupled from the Hubble expansion and undergo
gravitational collapse.

\subsection{Gravitational equilibrium}

There is a notable body of literature on BEC-CDM halos in
equilibrium and their properties, see e.g.
Ref.\cite{sin,hu,alcubierre,LL,RS} to name a few, with more
references to follow below.

The stationarity ansatz $\psi(\mathbf{r},t) =
\psi_s(\mathbf{r})e^{-i\mu t/\hbar}$, where the GP chemical
potential $\mu$ is fixed by the DM particle number, leads to the
stationary form of equ.(\ref{gp}), or (\ref{fluid}-\ref{hd3}), in
which the mass density $\rho = m|\psi_s|^2$ and, hence, the
gravitational potential $\Phi$ are \textit{time-independent}. Such
systems can be equivalently described by the GP energy functional,
which is given by
 \begin{equation} \label{energie}
   \mathcal{E}[\psi_s] = \int_V \left[\frac{\hbar^2}{2m}
 |\nabla \psi_s|^2 + \frac{m}{2}\Phi |\psi_s|^2 +
 \frac{g}{2}|\psi_s|^4\right]d^3\mathbf{r}.
 \end{equation}
Inserting again the decomposition $\psi_s(\mathbf{r}) =
|\psi_s|(\mathbf{r})e^{iS_s(\mathbf{r})}$ into (\ref{energie}) (and omitting the subscript 's'), the
total energy can be written as
  \begin{equation} \label{sumenerg}
   E = K + W + U_{SI},
    \end{equation}
 with the total kinetic energy
  \begin{equation} \label{kname}
   K \equiv \int_V \frac{\hbar^2}{2m}|\nabla \psi|^2 d^3\mathbf{r}
    = \int_V
   \frac{\hbar^2}{2m^2}(\nabla \sqrt{\rho})^2d^3\mathbf{r} + \int_V
   \frac{\rho}{2}\mathbf{v}^2d^3\mathbf{r},
    \end{equation}
 the gravitational potential energy
  \begin{equation} \label{wname}
   W \equiv \int_V \rho \Phi/2 ~d^3\mathbf{r},
    \end{equation}
 and the internal energy
  \begin{equation} \label{internal}
   U_{SI} \equiv \int_V g \rho^2/(2m^2) ~d^3\mathbf{r},
    \end{equation}
 which stems from the particle interactions, and which we have defined essentially as
 $U_{SI} = \int P_{SI} dV$ with
 $P_{SI}$ in (\ref{selfpressure}).

  These energy
 contributions enter the scalar virial theorem of an \textit{isolated} BEC-CDM halo under self-gravity, which reads as
  \begin{equation} \label{virial}
 2K + W + 3U_{SI} = 0.
  \end{equation}
As in classical gas dynamics, (\ref{virial}) (and possible boundary
terms) can be derived by multiplying the equations of motion in
fluid form, equ.(\ref{fluid}), by $\mathbf{r}$ and integrating the
resulting equation over volumes which enclose the system of interest. For an isolated body, a derivation involving a
scaling argument was presented in Ref.\cite{wang}.

Now, the size of an object in hydrostatic equilibrium can be determined by solving Equ.(\ref{gp})-(\ref{poisson}), or 
(\ref{poisson}) with
Equ.(\ref{fluid})-(\ref{hd3}). In fact, this has been done in Ref.\cite{MPS} in the limit where $g=0$, i.e. $P_{SI} = 0$, and
only quantum pressure will oppose gravity. The solution has no compact support, but the radius which includes 99 per cent of the mass
reads
 \begin{equation} \label{fuzzysize}
   R_{99} = 9.9 \hbar^2/(G M m^2).
    \end{equation}
It is easy to see that this is proportional to $\lambda_{deB}$,
Equ.(\ref{deB}), for a halo with corresponding virial velocity $v$.
All structure below $R_{99}$ will be suppressed by means of the
Heisenberg uncertainty principle. This regime has been called 'fuzzy DM' in Ref.\cite{hu}. Here, we will call this regime TYPE I
BEC-CDM. On the other hand, in the opposite regime, when
self-interaction is dominant, we can neglect $Q$ in
Equ.(\ref{fluid}). In that case, the equation of state is
an $(n=1)$-polytrope with corresponding radius
\begin{equation} \label{onesphere}
   R_0 = \pi
   \sqrt{\frac{g}{4\pi G m^2}},
    \end{equation}
   (see also Ref.\cite{goodman,BH} for complex SFDM and Ref.\cite{peebles,RT} for real SFDM). We will call this regime TYPE II BEC-CDM. Here,
$R_0$ is much larger than the corresponding value for
$\lambda_{deB}$, and, yet, it is the energy-independent cross
section in the quantum-mechanical low-energy limit,
Equ.(\ref{sigma}), which provides that scale via (\ref{coupling}) and (\ref{onesphere}). In either case, the
central DM densities of the object turn out to be lower, as compared
to CDM (see also Section 4.1). However, in both cases, the
hydrostatic equilibrium size does \textit{not} increase with the
mass $M$ of the object, and hence we can not limit ourselves to
either regime, in order to build up halos of variable size. The
considerations in the next sections address attempts to overcome
this problem, and in the course of that valuable constraints have
been found.

\section{New bounds on the SFDM particle parameters}

\subsection{Boson scattering and relaxation times}

We have seen that the hydrostatic equilibrium size of TYPE II
BEC-CDM halos is independent of halo mass, Equ.(\ref{onesphere}).
Therefore, in order to build a hierarchy of halos which resembles
observations, it is mandatory to go beyond the limitations of the
pure TYPE II limit. The authors of Ref.\cite{SG} have considered an
interesting scenario in which a pure BEC-CDM halo core of polytrope
radius (\ref{onesphere}) is enshrouded by an isothermal sphere of
bosons in thermal equilibrium. This isothermal envelope serves as
the halo outskirts and models flat galactic velocity profiles. In
the process of detailing the features of this model, however, the
authors can show that such a configuration must be ruled out, since
the compliance to two critical observations leads to contradictory
bounds on the boson parameters. As the authors of Ref.\cite{SG}
show, reproducing realistic velocity profiles requires a smooth
transition in the density between the polytropic core and the
isothermal envelope, which can be recast in a lower bound on the
boson mass. They find
\begin{displaymath}
m \geq 10 \times
\left(\frac{v_{c,\infty}}{100~\rm{km}~\rm{sec}^{-1}}\right)^{-1/4}\left(\frac{r_c}{1~\rm{kpc}}\right)^{-1/2}~\rm{eV}/c^2,
\end{displaymath}
where $v_{c,\infty}$ is the asymptotic circular velocity of the
isothermal envelope and $r_c$ is the minimum size of a halo core
supported only by particle repulsion, i.e. it follows Equ.(\ref{onesphere}).
Now, the authors also observe that, in the TYPE II regime, the DM scattering cross section per
unit mass can be written as 
\begin{equation} \label{cross}
\sigma_s/m = 8 G^2 R_0^4 m^5/(\pi^3 \hbar^4),
\end{equation}
by combining Equ.(\ref{sigma}), (\ref{coupling}) and (\ref{onesphere}).
Then, it is argued in Ref.\cite{SG} that upper bounds on $\sigma/m$
for the elastic-scattering particles in the self-interacting DM
model, i.e. CDM endowed with a finite cross section (referred to as ``SIDM'' in the literature), 
based upon comparing that model to astronomical
observations, should apply to BEC-CDM, as well\footnote{This
reasoning has also been applied in Ref.\cite{MUL}.}. The
interpretation of the Bullet cluster observations, for example, as a
nearly collisionless merger of two cluster-sized halos has been
found to limit $\sigma/m$ for SIDM halos to $(\sigma/m)_{max} <
1.25$ cm$^2$/g, according to Ref.\cite{randall}, and this limit is
imposed on Equ.(\ref{cross}) in Ref.\cite{SG} in order to arrive at
an upper bound on the boson mass of
\begin{equation} \label{SGupper}
m < 9.6 \times 10^{-4} \left(\frac{r_c}{1~\rm{kpc}}\right)^{-4/5}
~\rm{eV}/c^2.
\end{equation}
Obviously, these upper and lower bounds contradict each other for
any realistic choice of halo parameters, and so the isothermal
envelope can not make a cure to the radius-mass relationship of
BEC-CDM halos. We can see that this conclusion is valid,
even if we disregard the above SIDM limit from the Bullet cluster, as follows.
The relaxation time for achieving thermodynamic equilibrium
of a condensate fulfills to a good approximation the relationship,
  $\tau \simeq 1/(\sqrt{2}n \sigma_s \bar v)$
  with $n$ the condensate number density, $\sigma_s$ in Equ.
   (\ref{sigma}) and $\bar v$ the mean value of the velocity
 distribution of the particles (see Ref.\cite{GNZ}), where we use the non-relativistic description, appropriate
 for the matter-dominated epoch, and sufficient for the sake of our estimate. Re-writing this formula in terms of $\bar \rho
 \sigma_s/m$, we can write the
 relaxation time of a spherical, uniform halo (core) as
  \begin{displaymath}
   \tau
  = 1.889 \cdot
  10^{111}  \left(\frac{m}{m_H}\right)^{-5}\left(\frac{R}{1
  ~\rm{kpc}}\right)^{-3/2}\left(\frac{M}{10^{8} M_{\odot}}\right)^{5/2} \times
 \end{displaymath} 
   \begin{equation} \label{reltime}
  \times
  \left(\frac{\bar \rho}{\rm{GeV}/(c^2\rm{cm}^3)}\right)^{-1}
\left(\frac{\bar v}{100~\rm{km}/\rm{sec}}\right)^{-1} \rm{sec},
 \end{equation}
  where we define the characteristic particle mass
  \begin{equation} \label{mfidu}
   m_H = 1.066 \cdot 10^{-22}\left(\frac{R}{1 ~\rm{kpc}}\right)^{-1/2}\left(\frac{M}{10^{8}~ M_{\odot}}\right)^{-1/2}
  \rm{eV}/c^2
   \end{equation}
   (see Ref.\cite{RS2}), and $\bar \rho$ denotes the mean halo density. Choosing the size of a typical dwarf spheroidal galaxy with
   $R = 1$ kpc and $M = 10^8 M_{\odot}$, we see that $\tau$ will be
  \textit{larger} than a Hubble time,
  $\tau \gtrsim 10^{17}$ sec, if
 \begin{equation} \label{mupper}
  m \leq 1.066 \cdot 10^{-3}~ \rm{eV}/c^2,
  \end{equation}
  in good agreement with the bound in Equ.(\ref{SGupper}) from Ref.\cite{SG}.
 Inserting those values into Equ.(\ref{cross}) results in the
 corresponding upper bound of
 \begin{equation}
 \sigma_s/m \leq 2.1 ~\rm{cm}^2/\rm{g},
 \end{equation}
 hence not so different from the bound in Ref.\cite{randall}.  Thermodynamic
equilibrium is thus \textit{not} achieved for boson masses which
obey inequality (\ref{mupper}), i.e. for boson masses in which we
are interested\footnote{We note that these calculations differ from
that in Ref.\cite{RT}, where a self-annihilating real scalar field
is allowed to condense in the process of halo virialization, and a
Bose-enhancement factor enters their expression for the relaxation
time. This factor is only important at the BEC transition, but not
far below it, see also Ref.\cite{GNZ}. In the model we consider, however, condensation
happened in the very early Universe. Also, we do not agree on the
statement in Ref.\cite{RT} that it is impossible for complex SFDM to
fulfill cosmological constraints, along with the required central
density cusps of galaxies. In fact, density cusps are not supported
in complex SFDM either, and so boson self-annihilation is not a
prerequisite to explain different observations. See also Section
4.}. In Section 4.1, we will discuss a different scenario to
overcome the problem of the mass-independent size.

\subsection{SFDM as an extra relativistic degree of freedom
in the early Universe}

In the early Universe, the self-interaction term will dominate over
the mass term in Equ.(\ref{lag}). It can be shown that the
(background) equation of state of SFDM is then radiation-like. The timing of the transition from the
radiation-like to the dust-like phase of SFDM must be in accordance
with measurements of the redshift of matter-radiation equality
$z_{eq}$, as determined by the cosmic microwave background (CMB).
This requirement has been noted before in Refs.\cite{goodman,peebles,ALS}, and further investigated in Ref.\cite{LRS}. The
requirement that SFDM has fully morphed into a CDM-like fluid by the
time of $z_{eq}$ sets a constraint only on the ratio $g/(mc^2)^2$.
In Ref.\cite{LRS}, we found that\footnote{This is the value which
would make the equation of state parameter, $\langle \bar w \rangle
\equiv \langle \bar p \rangle/ \langle \bar \rho \rangle = 0.001$.
Note that Ref.\cite{LRS} uses the notation $\lambda$ for the
coupling strength $g$.}
\begin{equation} \label{zeq}
g/(mc^2)^2 \leq 4 \times 10^{-17}~ \rm{eV}^{-1} \rm{cm}^3.
\end{equation}
In addition, it has been found in Ref.\cite{LRS} that, for complex
SFDM, there is a transition at an even earlier epoch, when the
kinetic term due to the phase in Equ.(\ref{lag}) takes over and the equation of state of
complex SFDM changes from radiation-like to stiff-like, i.e. $\bar p
\simeq \bar \rho$. Once in the stiff phase, the expansion rate of
the universe is higher than in the radiation-dominated epoch, with
$H \propto a^{-3}$, instead of $H \propto a^{-2}$. Hence, complex
SFDM is the dominant cosmic component at these early epochs. In
contrast to Ref.\cite{ALS}, we do allow for SFDM-domination (in
its stiff phase), \textit{prior} to radiation-domination (see also
Ref.\cite{arbey}). The timely later transition from the stiff to the
radiation-like phase of SFDM, which is equivalent to SFDM-domination
giving way to radiation-domination, is constrained by the allowed
amount of relativistic degrees of freedom $N_{\rm{eff}}$ during Big
Bang nucleosynthesis (BBN). Requiring that this transition is
completed by the time of light nuclei production, we derive the
following constraints on the boson parameters in Ref.\cite{LRS},
\begin{equation} \label{massbound}
m \geq 2.4 \times 10^{-21} ~ \rm{eV}/c^2
\end{equation}
and
\begin{equation} \label{gbounds}
9.5 \times 10^{-19} \rm{eV}^{-1}\rm{cm}^3 \leq g/(mc^2)^2 \leq 1.5
\times 10^{-16} \rm{eV}^{-1}\rm{cm}^3.
\end{equation}
To derive these bounds, we imposed the (conservative) constraint
that the $N_{\rm{eff}}$ during BBN be all the time within $1\sigma$ of
the measured value, $N_{\rm{eff}}=3.71^{+0.47}_{-0.45}$, which we
adopt from Ref.\cite{steigman}. Note that the bounds in
Equ.(\ref{gbounds}) disfavor SFDM without self-interaction, i.e.
fuzzy DM or TYPE I BEC-CDM ! Now, combining the bounds on
$g/(mc^2)^2$ from equ.(\ref{zeq}) and (\ref{gbounds}), results in
corresponding bounds on the size of a virialized halo in TYPE II,
by means of Equ.(\ref{onesphere}), according to
\begin{equation} \label{sizebounds}
0.75 ~ \rm{kpc} \leq R_0 \leq 5.2 ~\rm{kpc}.
\end{equation}
As we pointed out in Ref.\cite{LRS}, it is a surprising and
curious fact that the constraint on the number of relativistic
degrees of freedom in the early Universe for the SFDM model leads to
bounds on the size scale, which fit so well into the expected
range of dwarf spheroidal and possibly smaller galaxies. By the same token, it is also clear
that these bounds can shift upon future measurements using BBN and CMB.

In light of these new bounds on the particle parameters of SFDM (and
necessarily BEC-CDM), as presented in this section, we will re-assess previous findings and study
the implications of this dark matter model on halo properties in the
next section.

\section{Implications for galactic structure}

\subsection{A hierarchy of BEC-CDM halos}

The characteristic length scales due to the quantum nature of
BEC-CDM result in the suppression of the formation of objects below
those scales, while structure formation on larger scales is expected
to follow the lore of standard CDM. Indeed, the power spectrum of
linear DM perturbations resembles the one for $\Lambda$CDM, except for the fact that the turnover
happens at lower wavenumbers and perturbations beyond a certain cutoff are suppressed, by analogy with neutrinos (hot DM)
or sterile neutrinos (warm DM). This
cutoff necessarily depends on the values of the SFDM particle
parameters, see Ref.\cite{hu} and Ref.\cite{MU}. Some model
parameters, for instance a real field with the popular value of $m
= 10^{-23}$ eV/$c^2$ and no self-interaction, i.e. fuzzy DM, are claimed to pass
the test in the sense of resembling $\Lambda$CDM on large scales, while appropriately suppressing structure
on small scales. On the other hand, we have seen above that models of complex SFDM without
self-interaction are disfavored for the current number of
$N_{\rm{eff}}$ during BBN. Also, the lower bound on particle mass derived in Ref.\cite{LRS}, Eq.(\ref{massbound}),
excludes masses as low as $10^{-23}$ eV$/c^2$. A much more detailed exploration of the
parameter space of SFDM models is necessary to draw further
conclusions. The study of the linear growth of structure in a
Universe with complex SFDM is more complicated because of the
additional degree of freedom due to the phase of the complex field.
Anisotropies in the corresponding perturbed energy-momentum tensor
have to be considered carefully, along with non-trivial boundary and
initial conditions. These issues are especially critical before the
matter-dominated epoch. A detailed analysis of perturbations in the
regime of matter-domination, however, has been presented in
Ref.\cite{Harko} and Ref.\cite{Chavanis}. The results of these
studies basically confirm that SFDM of the form of Equ.(\ref{fluid})
with (\ref{selfpressure}) has a growing mode solution for the linear
overdensity as for cold dust, i.e. $\delta \equiv \delta \rho/\bar
\rho \propto a$, with the scale factor $a$.

However, for lack of cosmological simulations, it shall be
sufficient here to outline the envisaged picture in an analytic
manner, continuing along the line of Ref.\cite{RS2}, by using a
simple top-hat collapse scenario (see also Ref.\cite{GU1} for SFDM
without self-interaction). A numerical infall study with
cosmological boundary conditions will be presented elsewhere. We
have noted in Section 2.2 that the equilibrium size $R$ of
virialized BEC-CDM halos of mass $M$ is in conflict with
observations: for a given set of particle parameters, the product
$RM = const.$ for TYPE I BEC-CDM, while $R$ is a function of density
(but independent of $M$) for TYPE II BEC-CDM, see
Equs.(\ref{fuzzysize}) and (\ref{onesphere}). In conjunction with
the disfavoring of TYPE I according to Ref.\cite{LRS}, this suggests
that it is desirable to improve upon models of TYPE II. The attempt
to enshroud a TYPE II halo core with an isothermal sphere of
thermalised bosons to overcome the size problem has been shown in
Ref.\cite{SG} to lead to contradictions, ruling out this scenario
(see Section 3.1). In Ref.\cite{RS2}, on the other hand, we have
argued that it is the kinetic energy of (possibly) coherent wave
motion, which may help to grow halos of the size range observed.
Numerical studies of the virialization process of isolated,
self-gravitating BEC blobs upon collapse show oscillations and mass
ejection away from the 'to-be' virialized core, termed
``gravitational cooling'' in Ref.\cite{SS} in the context of the
related phenomenon of boson stars. This feature is independent of
the particle masses chosen, as long as the system itself can be
described by a Schr\"odinger equation, see Ref.\cite{SS1,BSS,GU}. It
is therefore conceivable to envisage a picture where it is only the
minimum size halos (or halo cores in larger galaxies) which obey the
virial equilibrium of TYPE II BEC-CDM in the form of
Equ.(\ref{onesphere}), while larger halos have to have additional
energy contributions due to wave motion and possibly boundary terms
in the virial theorem.

Consider a 'classic' top-hat collapse scenario in an
Einstein-de-Sitter universe for BEC-CDM without kinetic energy to
derive the minimum size for BEC-CDM halos. We denote its quantities
with subscript 'zero'. A density perturbation is considered to
decouple from the general Hubble expansion at the turn-around radius
$r_{ta,0}$. There, we require that the total energy is entirely
gravitationally, i.e. $E_{ta,0} = W_{ta,0}$, and $U_{ta,0}=0,
K_{ta,0}=0$, see Equ.(\ref{sumenerg}). While $U_{ta,0}$ cannot
completely vanish in reality due to the finite (albeit, small)
self-interaction, neglecting its contribution is basically identical
to the requirement and expectation that BEC-CDM behaves dust-like,
i.e. like collisionless CDM, at the time of infall, prior to
virialization. According to the standard uniform sphere
approximation for the post-collapse, virialized object that results
from top-hat collapse, we assume the post-collapse sphere has
uniform density $\rho_0$, which fulfills virial equilibrium,
Equ.(\ref{virial}) with $K=0$, resulting in a corresponding virial
radius\footnote{The different prefactor compared to the one in
Equ.(\ref{onesphere}) stems from the fact that the density is
uniform, instead of following an $(n=1)$-polytropic run.} of
\begin{equation} \label{zerosphere}
  R_{TH,0} = \sqrt{\frac{15}{2}}\left(\frac{g}{4\pi
  G m^2}\right)^{1/2}.
   \end{equation}
It is easy to show that the collapse ratios are given by
 \begin{equation} \label{thratio}
  R_{TH,0}/r_{ta,0} = 2/3 ~~\mbox{ and }~~
 \rho_0/\rho_{ta,0} = (3/2)^3
  \end{equation}
(see Ref.\cite{RS2}), where $r_{ta,0}$ and $\rho_{ta,0}$ are the
radius and density of the pre-collapse sphere. Requiring in addition
that the top-hat density is proportional to the background density
at the time of collapse $\rho_{b,coll}$, we can see that
 \begin{equation} \label{THdensity}
 \rho_0 = \left(\frac{3}{2}\right)^3\rho_{ta,0} = \left(\frac{3}{2}\right)^3
 \frac{9\pi^2}{4}\rho_{b,coll} = \frac{243\pi^2}{32}\rho_{b,coll}
 \simeq 75 ~\rho_{b,coll}.
 \end{equation}
  Note that this factor is significantly smaller than the standard value of
  $18\pi^2 \simeq 178$ for collisionless CDM. Thus, already at this simple
  level is it evident that BEC-CDM will result in collapsed structures of lower
  density\footnote{This suggests that a too literal adoption of the
  standard CDM framework for halo mass functions and profiles to dark matter models with a significant fraction of
  ultra-light axions can be problematic, see Ref.\cite{MS}. It is very interesting,
  though, that the analysis of these authors also hint to a mass of $m \sim 10^{-21}$
  eV$/c^2$, close to the bound in Equ.(\ref{massbound}), by using
  entirely different constraints.}. However, $R_{TH,0}$ and $r_{ta,0}$ have a unique value,
  once the ratio of particle parameters $g/m^2$ is fixed. In practice, we will want
  to fix the minimum size $R_{TH,0}$, motivated by galaxy
  observations, to determine the allowed value of particle parameters.
  The exact value of $R_{TH,0}$, however, is not necessary in outlining the general idea.

Now, in order to build halos of size $R$ \textit{larger} than
$R_{TH,0}$, we include an (effective) kinetic term $K_{\rm{eff}}$ in
the virial theorem (\ref{virial}), which shall capture the overall
kinetic energy due to wave motions. For the general argument
outlined here, it is sufficient to consider a non-vanishing bulk
velocity $\mathbf{v}$, such that $K_{\rm{eff}} = \int \frac{\rho}{2}
\mathbf{v}^2 dV > 0$. Since we restrict our analysis to top-hats
with uniform density $\rho$, we may consider the gross average of
$K_{\rm{eff}} = \frac{3}{2} M \sigma_v^2$ with the top-hat mass $M$
and velocity dispersion $\sigma_v^2 = \frac{1}{3}\langle
\mathbf{v}^2\rangle$. Using (\ref{virial}) with that form of
$K_{\rm{eff}}$, we can write the velocity dispersion as
 \begin{equation}
 \sigma_v^2 = G M/(5R) - K\rho = G M/(5R)[1 - (R_{TH,0}/R)]^2,
 \end{equation}
  or in fiducial units (after taking the root)
\begin{equation} \label{VD}
 \sigma_v = 9.275 \left[1 - \left(\frac{R_{TH,0}}{R}\right)^2\right]^{1/2}\left(\frac{M}{10^8 M_{\odot}}\right)^{1/2}\left(\frac{1
 ~\rm{kpc}}{R}\right)^{1/2} \rm{km}/\rm{sec}.
 \end{equation}
  Additionally,
in order for the virial radius $R$ to depend on halo mass and
redshift of collapse $z_{coll}$, as they do for standard CDM, we
require the top-hat density to be a fixed fraction of the background
density at $z_{coll}$, that is, $\rho = C \rho_{b, coll}$, where the factor $C$
has to be determined numerically using Equ.(\ref{collapseratio2})
below. The corresponding total mass is then $M = 4\pi R^3 C
  \rho_{b, coll}/3$, which yields a radius
   \begin{equation} \label{RM}
  R(M,z_{coll}) = [3M/(4\pi
  C \rho_{b, coll})]^{1/3}.
  \end{equation}
Repeating the above calculation for these \textit{larger} top-hats
with $E_{ta} = W_{ta}$ and $E_{post-collapse} = W + U + K_{\rm{eff}}$, we
obtain for the corresponding collapse ratio
 \begin{equation} \label{collapseratio1}
\frac{R}{r_{ta}} = 1 - \frac{\sigma_v^2}{2(\sigma_v^2 + K\rho)} -
\frac{K \rho}{3(\sigma_v^2 + K\rho)} = \frac{\sigma_v^2/2 +
2K\rho/3}{\sigma_v^2 + K\rho}.
\end{equation}
The corresponding generalization of Equ.(\ref{THdensity}) is then
given by
\begin{equation} \label{collapseratio2}
\rho = \frac{9\pi^2}{4}\rho_{b,coll}\left(\frac{\sigma_v^2 +
K\rho}{\sigma_v^2/2 + 2K\rho/3}\right)^3,
\end{equation}
again assuming an EdS universe with dust in order to relate
$\rho_{ta} = 9\pi^2 \rho_{b,coll}/4$. However, note that Equ.(\ref{collapseratio1}) and (\ref{collapseratio2}) are
not explicit and must be solved numerically for any given $K$ and
non-vanishing $\sigma_v$. 
The standard CDM results of $\sigma_v = \sqrt{GM/(5R)}$, $R/r_{ta} =
1/2$ and $\rho/\rho_{b,coll}=18\pi^2$ are recovered if $g/m^2 \equiv
0$, since then $K = 0$ and hence $R_{TH,0} = 0$. So, we see that the
collapse ratios $R/r_{ta}$ and $\rho/\rho_{b,coll}$ are \textit{not}
universal due to the lower size cutoff of DM perturbations, provided
by BEC-CDM. The resulting smallest halos have a size of $R_{TH,0}$,
a mass of $M_{min} = 4\pi \rho_0 R_{TH,0}^3/3$ and $\sigma_v=0$ by
construction. On the other hand, larger halos will follow the
relationship (\ref{RM}) with a non-vanishing velocity dispersion due
to internal wave motion, according to (\ref{VD}). This guarantees
that BEC-CDM halos of mass $M > M_{min}$ would share the mass-radius
relation of halos in the standard CDM model, if halos of a given mass $M$ typically
  collapse at the same time as they do for standard CDM.
  The standard CDM relationships will be
more and more accurate the higher the velocity dispersion of the DM
halo becomes, i.e. the larger $R/R_{TH,0}$. Therefore, the
clustering and halo properties on scales beyond the smallest
galaxies and halo cores, respectively, will be more or less the same
for BEC-CDM as for collisionless CDM.

The particle parameters, which enter $R_{TH,0}$ in the combination
$g/m^2$ may now be chosen such that $M_{min}$ corresponds to the halo mass of the 
smallest observed galaxies, as well as to the DM core mass of large galaxies.
To address the cusp-core issue, along with
  the ``missing satellite'' problem, we may choose a
  fiducial value of $R_{TH,0} = 1 ~\rm{kpc}$, the order of magnitude
  at which these problems arise, and which is in accordance with the
  bounds in Equ.(\ref{sizebounds}). Also, observations of Milky
  Way dwarf spheroidal galaxies suggest that they host about $10^7
  ~M_{\odot}$ within the central 300 pc Ref.\cite{Strigari},
  $(2-7)\times 10^7 ~M_{\odot}$ within about 600 pc Ref.\cite{Walker},
  and as a result virial masses of $10^8-10^9~M_{\odot}$ with maximum velocity dispersions of order $\sigma_v \simeq 10$ km/sec. Thus,
  a fiducial choice of $R_{TH,0} = 1$ kpc
  and $M_{min} = 10^8~M_{\odot}$ seems reasonable.

\subsection{Signature effects unique to BEC-CDM halos}

\subsubsection{Halo shapes and quantized vortices}

The standard scenario of structure formation expects that halos will
acquire angular momentum in the course of tidal-torquing due to the
surrounding large-scale structure. Galaxies are observed to have
angular momentum, which must have been seeded by the DM in whose
potential wells they formed. N-body simulations of cosmic DM
structure formation do indeed confirm the basic expectations from
tidal-torque theory, confirming non-spherical shapes, even though
the amount of angular momentum is small, and DM halos are far from
being rotationally supported. Values for the dimensionless spin
parameter $\lambda \equiv L |E|^{1/2}/(GM^{5/2})$,
  with $L$ the
total angular momentum, and $E$ the total energy of a halo of mass
$M$, cover a typical range of about $0.01 - 0.1$ with median values
of about $0.03-0.05$, for halos in the present Universe
Ref.\cite{BE,GY}, but also for primordial halos Ref.\cite{ON,SCMF}.
Standard CDM also seems to prefer a distribution of shapes, with
oblate and prolate axis ratios for higher and highest mass halos.
The study of halo shapes in BEC-CDM, on the other hand, has hardly
been considered and results are very scarce. In Ref.\cite{RS}, we
have studied some simple, but analytic models. BEC-CDM halo velocity
fields obey the irrotationality condition for angular momenta which
are lower than the threshold for forming a (singly-quantized)
vortex, which breaks this condition locally. For an axisymmetric
vortex in the center of a halo, this threshold is given by $L_{QM}
\equiv \frac{M}{m} \hbar$, which, for a given $\lambda$ and $L$
becomes a minimum condition on the particle mass. Rotation of the
halo imprints a non-vanishing phase gradient in the scalar field,
and we expect non-spherical shapes. In Ref.\cite{RS}, we have shown
that the irrotationality condition will force TYPE II BEC-CDM halos
to be of prolate form. More precisely, it can be shown that halos
can be described by Riemann-S ellipsoids before and after vortex
formation. Recently, the authors of Ref.\cite{Getal} conclude from
numerical tests that the inclusion of angular momentum may be a way
to make rotation curves of BEC-CDM halos fit better to observations,
at least as long as no vortices arise.

The formation of quantized vortices in SFDM has attracted
comparatively more attention. In Ref.\cite{SM}, it was argued that,
if DM is composed of ultra-light BEC-CDM, the rotation velocity of
the Andromeda galaxy would be sufficient to create vortices.
Subsequently, it was shown in Ref.\cite{YM} that a lattice of about
500 vortices can produce a velocity profile which can fit data for
the Milky Way. Recently, in Ref.\cite{KL} the detailed density
profile of a spherical halo in the presence of a central vortex was
calculated, and limits on the boson parameters for a vortex to form
in the Andromeda galaxy were derived. All those works, however,
were limited to spherical halo shapes. On the other hand, in
Ref.\cite{RS} we have derived the bounds on the boson parameters for
vortex formation, resulting from self-consistent Maclaurin and
Riemann-S ellipsoidal solutions of the GPP system, which have the
same values for the spin parameter $\lambda$, as we expect from
standard CDM. Vortex formation requires a minimum mass $m \geq
m_{crit}$ and a minimum coupling strength $g \geq g_{crit}$. These
critical values are smaller for larger $\lambda$, from which larger
vortex cores result. According to Table II and III of Ref.\cite{RS2}
for a fiducial halo of dwarf spheroidal size of $R=1$ kpc, the critical values for the particle parameters above
which vortex formation happens are \textit{lower} than the bound in
Equ.(\ref{massbound}) and (\ref{gbounds}) \textit{if} $\lambda \geq 0.05$ or $\lambda
\geq 0.1$, respectively, depending on the halo model. That is,
vortex formation can be avoided only if $\lambda \leq 0.05$. This is
an interesting result, since it can imply a \textit{depleted} DM
density in the centers of such galaxies due to a central vortex. More detailed modeling
along with a careful comparison to observed velocity profiles of
dwarf spheroidals will be an important additional test-bed for SFDM.

\subsubsection{Halo mergers}

On the small scales on which BEC-CDM differs from CDM, as described
above, the wave coherence of BEC-CDM can result in distinctive
features upon halo collision and merging. Indeed, for different
potentials of real scalar fields, it has been shown in previous
works, see e.g. Ref.\cite{GG,VM}, that two equal-sized
self-gravitating BEC-CDM blobs will be colliding and merging to form
a single structure if the total energy of the system is negative.
The relaxation process results again in the ejection of scalar
field, i.e. gravitational cooling. However, if the total energy of
the system is positive, the two blobs can pass through each other
intactly, i.e. they can exhibit solitonic behavior. While
previous studies have established the generality of these phenomena,
we still lack a detailed modeling of halo mergers and comparison to
observations (for instance to the Bullet galaxy cluster), which
would result in particle parameter space exclusion regions. It
remains to be seen whether the allowed parameter space of SFDM from
cosmological and galactic observations may be challengend further by
the distinctive phenomenology of halo mergers. Since BEC-CDM halos
become more and more CDM-like at higher masses, we would not expect
dramatic differences on the scales of galaxy cluster collisions,
except for the very cluster centers of order 1 kpc. As far as
observations can identify the degree of collisionality on those
scales, whether for clusters or galaxies, this will enable to
constrain SFDM (and hence BEC-CDM) further. However, cosmological
simulations of SFDM are needed to answer those questions in more
detail. In fact, the work of Ref.\cite{woo} remains the only realistic structure formation simulation of
complex SFDM, but without self-interaction (i.e. fuzzy DM). Indeed, a few halo
mergers with rather complicated interference patters are observed
within a simulation volume of side length 1 Mpc/h. However, one
unexpected outcome consists of cusps in the density profiles of
those halos. This is very counter-intuitive, given the analytial
expectation that cusps are prohibited due to the inherent
characteristic length scales of BEC-CDM, and is also in contrast to
numerical results of Ref.\cite{hu} in 1D. We believe that the use of
the pseudo-spectral method in solving the Schr\"odinger equation in
Ref.\cite{woo} could be too insensitive to capture the detailed
shock physics at halo formation and merging, and may lead to poor resolutions of the halo centers,
after all.

This result along with the other ones we have been trying to
summarize in this review will hopefully spur on more activity in
complex scalar field dark matter, the interesting alternative to
collisionless CDM and the WIMP paradigm.

\section*{Acknowledgments}

This work was supported in part by U.S. NSF grants AST-0708176,
AST-1009799 and NASA grants NNX07AH09G, NNG04G177G, NNX11AE09G to
PRS. TRD also acknowledges support by the Texas Cosmology Center of
the University of Texas at Austin and by the Michigan Center for Theoretical Physics of
the University of Michigan.


\begin{thebibliography}{0}


\bibitem{Peccei}
R.D. Peccei, in {\it Lect.Notes Phys.} {\bf 741}, 3-17, Springer,
2008

\bibitem{SY}
P.Sikivie, Q.Yang, {\it Phys.Rev.Lett.} {\bf 103}, 111301 (2009)


\bibitem{FHSW}
J.A. Frieman, C.T. Hill, A. Stebbins, I. Waga, {\it Phys.Rev.Lett.}
{\bf 75}, 2077 (1995)


\bibitem{axiverse}
   A. Arvanitaki, S. Dimopoulos, S. Dubovsky, N. Kaloper, J. March-Russell, {\it Phys.Rev. D} {\bf 81},
   123530 (2010)


\bibitem{GZ}
   U. G\"unther, A. Zhuk, {\it Phys.Rev. D} {\bf 56}, 6391 (1997)


\bibitem{FMT}
T. Fukuyama, M. Morikawa, T. Tatekawa, {\it J. Cosmol. Astropart. Phys.} {\bf 06}, 033
(2008)

\bibitem{FM}
T. Fukuyama, M. Morikawa, {\it Phys.Rev. D} {\bf 80}, 063520 (2009)


\bibitem{Carroll}
   S.M. Carroll, {\it Phys.Rev.Lett.} {\bf 81}, 3067 (1998)

\bibitem{FG}
F. Ferrer, J.A.Grifols, {\it J. Cosmol. Astropart. Phys.} {\bf 12}, 012 (2004)




\bibitem{Weinberg}
D.H. Weinberg, J.S. Bullock, F. Governato, R. Kuzio de Naray, A.H.G.
Peter, {\it arXiv:1306.0913}, (2013)


\bibitem{MM}
J. Maga\~na, T. Matos, {\it J.Phys.:Conf.Series} {\bf 378}, 012012
(2012)

\bibitem{Guzman}
F.S. Guzm\'an, {\it J.Phys.:Conf.Series} {\bf 91}, 012003 (2007)

\bibitem{madelung}
 E. Madelung, {\it Z.f\"ur Phys.} {\bf 40}, 322 (1927)



\bibitem{sin}
    S.J. Sin, {\it Phys.Rev. D} {\bf 50}, 3650 (1994)

\bibitem{hu}
    W. Hu, R. Barkana, A. Gruzinov, {\it Phys.Rev.Lett.} {\bf 85}, 1158
    (2000)

\bibitem{alcubierre}
  M. Alcubierre, F.S. Guzm\'an, T. Matos, D. N\'u\~nez,
  L.A. Ure\~na-L\'opez, P. Wiederhold, {\it Class.Quant.Grav.} {\bf 19}, 5017
  (2002)


\bibitem{LL}
     J.-W. Lee, S. Lim, {\it J. Cosmol. Astropart. Phys.} {\bf 7}, 01 (2010)


\bibitem{RS}
    T. Rindler-Daller, P.R. Shapiro, {\it Mon. Not. R. Astron. Soc.} {\bf 422}, 135 (2012)




\bibitem{wang}
    X.Z. Wang, {\it Phys.Rev. D} {\bf 64}, 124009 (2001)

\bibitem{MPS}
     M. Membrado, A.F. Pacheco, J. Sa\~nudo, {\it Phys.Rev.A} {\bf 39},
     4207 (1989)



\bibitem{goodman}
    J. Goodman, {\it New Astronomy} {\bf 5}, no.2, 103 (2000)


\bibitem{BH}
    C.G. B\"ohmer, T. Harko, {\it J. Cosmol. Astropart. Phys.} {\bf 06}, 025 (2007)


\bibitem{peebles}
     P.J.E. Peebles, {\it Ap. J.} {\bf 534}, L127 (2000)

\bibitem{RT}
A. Riotto, I. Tkachev, {\it Phys.Lett. B} {\bf 484}, 177 (2000)



\bibitem{SG}
    Z. Slepian, J. Goodman, {\it Mon. Not. R. Astron. Soc.} {\bf 427}, 839 (2012)


\bibitem{MUL}
T. Matos, L.A. Ure\~na-L\'opez, {\it Phys.Lett. B} {\bf 538}, 246
(2002)


\bibitem{randall}
S.W. Randall, M. Markevitch, D. Clowe, A.H. Gonzalez, M. Brada\v{c},
{\it Ap. J.} {\bf 679}, 1173 (2008)



\bibitem{GNZ}
A. Griffin, T. Nikuni, E. Zaremba, {\it Bose-condensed gases at
finite temperatures}, Cambridge Univ.Press (2009)



 \bibitem{RS2}
 T. Rindler-Daller, P.R. Shapiro,  {\it Astrophys. Space Sci. Proc.} \textbf{38} (2013), in press, arXiv:1209.1835


\bibitem{ALS}
   A. Arbey, J. Lesgourgues, P. Salati, {\it Phys.Rev. D} {\bf 65},
   083514 (2002)

\bibitem{LRS}
     B. Li, T. Rindler-Daller, P.R. Shapiro, arXiv:1310.6061 (2013)

\bibitem{arbey}
A. Arbey, {\it Phys.Rev. D} {\bf 74}, 043516 (2006)

\bibitem{steigman}
G. Steigman, arXiv:1208.0032 (2012)


\bibitem{MU}
     T. Matos, L.A. Ure\~na-L\'opez, {\it Phys.Rev. D} {\bf 63}, 063506
     (2001)



\bibitem{Harko}
T. Harko, {\it Mon. Not. R. Astron. Soc.} {\bf 413}, 3095 (2011)


\bibitem{Chavanis}
P.H. Chavanis, {\it Astron. Astrophys.} {\bf 537}, A127 (2012)


\bibitem{GU1}
    F.S. Guzm\'an, L.A. Ure\~na-L\'opez, {\it Phys.Rev. D} {\bf 68},
    024023 (2003)

\bibitem{SS}
E. Seidel, W.-M. Suen, {\it Phys.Rev.Lett.} {\bf 72}, 2516 (1994)

\bibitem{SS1}
E. Seidel, W.-M. Suen, {\it Phys.Rev. D} {\bf 42}, 384 (1990)

\bibitem{BSS}
J. Balakrishna, E. Seidel, W.-M. Suen, {\it Phys.Rev. D} {\bf 58},
104004 (1998)


\bibitem{GU}
    F.S. Guzm\'an, L.A. Ure\~na-L\'opez, {\it Phys.Rev. D} {\bf 69},
    124033 (2004)



\bibitem{MS}
D.J.E. Marsh, J. Silk, {\it Mon. Not. R. Astron. Soc.} {\bf 437}, 2652 (2014) 


\bibitem{Strigari}
L.E. Strigari, J.S. Bullock, M. Kaplinghat, J.D. Simon, M. Geha, B.
Willman, M.G. Walker, {\it Nature} {\bf 454}, 1096 (2008)


\bibitem{Walker}
M.G. Walker, M. Mateo, E.W. Olszewski, O.Y. Gnedin, X. Wang, B. Sen,
M. Woodroofe, {\it Ap. J.} {\bf 667}, L53 (2007)


\bibitem{BE}
    J. Barnes, G. Efstathiou, {\it Ap. J.} {\bf 319}, 575 (1987)

\bibitem{GY}
S. Gottl\"ober, G. Yepes, {\it Ap. J.} {\bf 664}, 117 (2007)


\bibitem{ON}
B.W. O'Shea, M.L. Norman, {\it Ap. J.} {\bf 654}, 66 (2007)

\bibitem{SCMF}
R.S. de Souza, B. Ciardi, U. Maio, A. Ferrara, {\it Mon. Not. R. Astron. Soc.} {\bf
428}, 2109 (2013)


\bibitem{Getal}
F.S. Guzm\'an, F.D. Lora-Clavijo, J.J. Gonz\'alez-Avil\'es, F.J.
Rivera-Paleo, arXiv:1310.3909


\bibitem{SM}
    M.P. Silverman, R.L. Mallet, {\it Gen.Rel.Grav.} {\bf 34}, 633
    (2002)


\bibitem{YM}
    R.P. Yu, M.J. Morgan, {\it Class.Quant.Grav.} {\bf 19}, L157 (2002)


\bibitem{KL}
    B. Kain, Y. Ling, {\it Phys.Rev. D} {\bf 82}, 064042 (2010)


\bibitem{GG}
J.A. Gonz\'alez, F.S. Guzm\'an, {\it Phys.Rev. D} {\bf 83}, 103513
(2011)


\bibitem{VM}
D. Casta\~neda Valle, E.W. Mielke, {\it Ann.of Phys.} {\bf 336}, 245
(2013)


\bibitem{woo}
    T.-P. Woo, T. Chiueh, {\it Ap. J.} {\bf 697}, 850 (2009)




\end{thebibliography}
\end{document}